\documentclass[12pt]{article} 
\usepackage[draft]{hyperref}
\usepackage{amsmath}
\usepackage{braket}

\usepackage{graphicx}
\newcommand{\address}[1]{\\\small\textit{#1}\\}

\newcommand{\ws}{\omega_s}
\newcommand{\wi}{\omega_i}
\newcommand{\wc}{\omega_c}
\newcommand{\dw}{\Delta\omega}

\newcommand{\degpoint}{\ensuremath{\left(\wc,\wc\right)}}
\newcommand{\dk}{\Delta\!k}
\newcommand{\dks}{\dk'_s}
\newcommand{\dki}{\dk'_i}
\newcommand{\func}[2]{\mathrm{#1}\!\left(#2\right)}
\newcommand{\prob}{\func{p}{\tau}}
\newcommand{\Real}[1]{\mathrm{Re}\left[#1\right]}
\newcommand{\Abs}[1]{\left|#1\right|}
\newcommand{\etothe}[1]{e^{#1}}
\newcommand{\gauss}[2]{\etothe{-\frac{\left(#1\right)^2}{2 #2^2}}}

\newcommand{\etothemfrac}[2]{e^{-\frac{#1}{#2}}}

\newcommand{\Sinc}[1]{\func{sinc}{#1}}
\newcommand{\Cos}[1]{\func{cos}{#1}}

\newcommand{\optelem}[2]{\(\mathrm{#1_{#2}}\)}
\newcommand{\qwp}[1]{\optelem{QWP}{#1}}

\newcommand{\pbs}[1]{\optelem{PBS}{#1}}
\newcommand{\sflt}[1]{\optelem{SF}{#1}}

\newcommand{\desc}[2]{\hat{#1}_{#2}}
\newcommand{\asc}[2]{\desc{#1}{#2}^{\dagger}}

\newcommand{\vacstate}{\ket{0}}
\newcommand{\spc}{\!\!}
\newcommand{\intd}{\mathrm{d}}
\newcommand{\intv}[1]{\int\spc\intd#1\,\,}
\newcommand{\intvv}[2]{\int\spc\int\!\!\intd#1 \intd#2\,\,}
\newcommand{\freqint}{\intvv\ws\wi}

\newcommand{\uval}[2]{\ensuremath{#1\,\mathrm{#2}}}

\begin{document}
\title{Broadband frequency mode entanglement in waveguided PDC}\maketitle
\author{Andreas Eckstein$^*$ and Christine Silberhorn}
\address{Max Planck Research Group, G\"unther-Scharowsky-Str. 1, 91054 Erlangen}
\address{$^*$Corresponding author: aeckstein@optik.uni-erlangen.de}

We report the observation of beatings of the coincidence event rate in a Hong-Ou-Mandel interference (HOMI) between signal and idler photons from a parametric downconversion (PDC) process inside a multi-mode KTP waveguide. As explanation we introduce bi-photonic states entangled in their broadband frequency modes generated by waveguide mode triples and propose a suitable entanglement detection scheme.\\\\



\noindent Waveguided parametric downconversion is a promising approach to realize photon pair sources for applications in quantum communication\cite{URen03} or quantum imaging\cite{Lugiato02}. The mode confinement of the waveguide allows for a significant increase of the  source brightness in comparison to bulk crystal setups\cite{Tanzilli01, Anderson95, URen04a}.
In addition, owing to the guiding structure, a PDC process in a nonlinear waveguide will produce photon pairs in distinct spatial modes\cite{Banaszek01}, which contrasts to the continuous distributions defined by phasematching inside bulk crystals. While for bulk crystals the photon pairs are generally correlated in their transverse momenta, the waveguide also enforces decorrelation of the spatial signal and idler modes. However, in general a PDC process introduces an interrelation between the spatial properties of the pump and the spectral distributions of signal and idler photons\cite{Banaszek01,Carrasco04} as well. We expect to find a similar behavior for the wave\-guided case.

The Hong-Ou-Mandel interference\cite{Mandel87} is widely used to probe the spatial-spectral structure of single photon pairs. It measures the exchange parity of a two-photon state by detecting a coincidence probability in dependence of the overlap of both input modes\cite{Branning00} and thus offers information on the correlations of a bi-photonic state. This has been applied to the polarization\cite{Michler96}, temporal\cite{Atature01} or spatial\cite{Nogueira01} degree of freedom: deviation from the classic dip signature hints at entanglement.

\begin{figure}
\begin{center}
\includegraphics[scale=1]{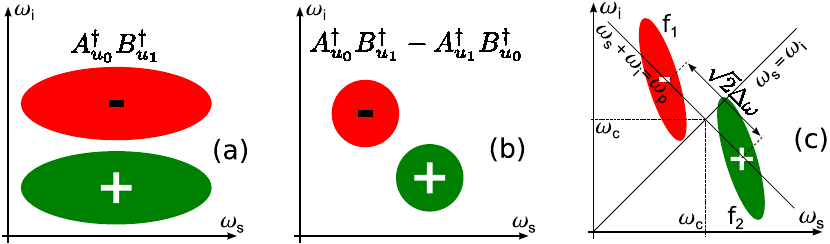}
\caption{(Color online) (a)\&(b) Shape of joint spectral amplitudes for combinations of lowest order Hermite functions, (c) Shape of superimposed PDC processes}\label{broadbandentangled}
\end{center}
\end{figure}

In this paper, we employ waveguided PDC and study the spectral structure of its resulting bi-photonic state, which is strongly affected by the discrete spatial modes of the pump. We introduce entanglement of discrete broadband frequency modes, and for the first time provide a characterization of states with disjoint biphoton spectral intensities as illustrated in Fig. \ref{broadbandentangled}(b). For such states, a HOM experiment results in beatings in the coincidence event detection, which have been reported for settings using frequency post-selection \cite{Mandel88,Rarity90,Kim03} or active phase manipulation \cite{Dayan07}. We on the other hand demonstrate generation of these states inside a waveguide without additional manipulation and use the signature of a HOM experiment as a witness of entanglement, analogous to the the well known Bell state analyzer\cite{Mann95,Michler96}.

Broadband photon creation \(\asc{A}{f}\) operators may be defined by monochromatic creation operators \(\func{\asc{a}{}}{\omega}\) as
\begin{align}
\asc{A}{f}=\int\!\mathrm{d}\omega\,\,\func{f}{\omega}\func{\asc{a}{}}{\omega}\label{broadbandmode}
\end{align}
with the function \(\func{f}{\omega}\) specifying the spectral amplitude of the photon.
Any amplitude function $f(\omega)$ can be expanded into a basis set of orthogonal functions, such that the single photon state may be decomposed into a superposition of broadband single photon states with orthonormal amplitude functions: \(\ket{\psi}=\asc{A}{f} \vacstate=\sum_{i=0}^{\infty}c_i\asc{A}{u_i}\vacstate\).
Let \(\func{u_i}{\omega}\) denote the  Hermite functions and let \(\asc{A}{u_i}\) be a broadband mode creation operator as defined in eq. \ref{broadbandmode} with a Hermite function as amplitude.

Let us now construct a maximally entangled two-photon state using the \(\func{u_i}{\omega}\) as a basis of frequency modes
\begin{align}
\ket{\psi^-}=\frac{1}{\sqrt{2}} \left[\asc{A}{u_0}\asc{B}{u_1}-\asc{A}{u_1}\asc{B}{u_0}\right]\vacstate\label{bellstate}
\end{align}
with A and B signifying signal and idler modes, respectively. The structure of \(\ket{\psi^-}\) is formally analogous to the polarization entangled singlet Bell-state, where the Hermite functions \(\func{u_0}{\omega}\) and  \(\func{u_1}{\omega}\) are replaced by the horizontal and vertical polarization. The resulting joint spectral amplitudes with contributions \(\func{u_i}{\omega_i}\) and \(\func{u_j}{\omega_s}\) for two-party systems are illustrated in Fig. \ref{broadbandentangled}(a) and (b).
A Bell state analyzer\cite{Mann95} allows to identify a polarization singlet state by HOM-type experiment by measuring quantum interference at a balanced beam splitter between the modes \(A\) and \(B\) in dependence of their  polarization overlap\cite{Michler96}. Here only an entangled singlet state exhibits the signature of photon anti-bunching, which is a proof of entanglement.

In the following we will show that an analogous criterion can be constructed for broadband entangled modes in a standard HOM experiment. Consider a standard HOM experiment for a bi-photonic state with joint spectral amplitude \(\func{f}{\ws,\wi}\). The two photons are fed into a balanced beamsplitter with varying time delay \(\tau\) and the  outputs are monitored by binary photon detectors. The probability to detect a coincidence event from both detectors for each arriving photon pair is then given by
\begin{align}
\prob=\frac{1}{2} - \frac{1}{2} \frac{\Real{\freqint\func{f^*}{\ws,\wi}\func{f}{\wi,\ws}\etothe{-\imath\tau\left(\ws-\wi\right)}}}{\freqint\left|\func{f}{\ws,\wi}\right|^2}\label{homgeneric}
\end{align}
For a separable input bi-photon state with factorisable joint spectral amplitude
\(\ket{\psi_s} \otimes \ket{\psi_i}=\freqint \func{f_1}{\ws}\func{f_2}{\wi} \func{\asc{a}{s}}{\ws}\func{\asc{a}{i}}{\wi}\vacstate\)
we find
\begin{align}
\prob=\frac{1}{2} - \frac{1}{2} \frac
{\Abs{\intv{\omega}\func{f_1^*}{\omega}\func{f_2}{\omega}\etothe{-\imath\tau\omega}}^2}
{\intv{\omega}\Abs{\func{f_1}{\omega}}^2\intv{\omega}\Abs{\func{f_2}{\omega}}^2}\label{homreal}
\end{align}
where positivity of the fraction term ensures \(\prob<0.5\). We conclude that \(p_c > 0.5\)~implies a non-separable, i.~e. entangled two-photon input state:
\begin{align}
\prob>0.5 \Rightarrow \ket{\psi_{s,i}} \neq \ket{\psi_s} \otimes \ket{\psi_i}
\end{align}
Thus the HOM coincidence detection probability \(\prob\) serves as an witness of entanglement in analogy to the polarization Bell state analyzer. We note that additional entanglement in other degrees of freedom will not invalidate this relation. In addition we would like to stress that in contrast to entanglement detection in other degrees of freedom, e.~g. polarization or transverse momenta, the presence of broadband frequency mode entanglement in a bi-photon with amplitude \(\func{f}{\ws,\wi}=\func{u_0}{\ws}\func{u_1}{\wi}-\func{u_1}{\ws}\func{u_0}{\wi}\) leads to a beating signature in \(\prob\) in general whereas all other types will just switch between Gaussian dip and ``bump'' shape.

\begin{figure}
\begin{center}
\includegraphics[scale=1]{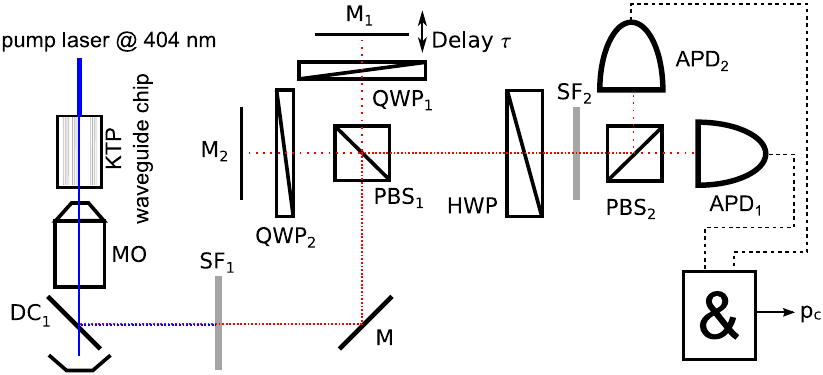}
\caption{(Color online) Experimental setup}\label{experiment}
\end{center}
\end{figure}
In the experiment as shown in Fig. \ref{experiment}, we employ a periodically poled KTP waveguide with dimensions \(\uval{12}{mm}\times\uval{4}{\mu m}\times\uval{4}{\mu m}\), which is pumped by a picosecond diode laser at \uval{404}{nm} with \uval{1.4}{nm} FWHM and a repetition rate of \uval{16}{MHz}. We excite a type II PDC process with orthogonally polarized signal and idler photons and central frequencies around \uval{808}{nm}. After the signal and idler photons are split at \pbs{1}, we utilize a Michelson type interferometer configuration with two additional quarter wave plates \qwp{1} and \qwp{2} to introduce a variable delay between the orthogonally polarized photons. A spectral filter \sflt{1} centered around \uval{808}{nm} with \uval{3}{nm} fwhm is introduced to enhance signal-to-noise ratio. We then enter a common path HOM interferometer of the first type\cite{Shih94}: the half wave plate HWP and \pbs{2} act as a balanced beam splitter on both incoming polarization modes where the HOM interference takes place. Subsequently, the output photons are detected with a pair of avalanche photo diodes (APDs), and coincident photons impinging on the detectors are counted separately. Single photon event count rates are of the order of 100 kHz or \(\frac{1}{160}\) events per pump pulse, thus justifying the neglection of multiple photon pair creation during one pump pulse.

\begin{figure}
\begin{center}
\begin{center}
\includegraphics[scale=1]{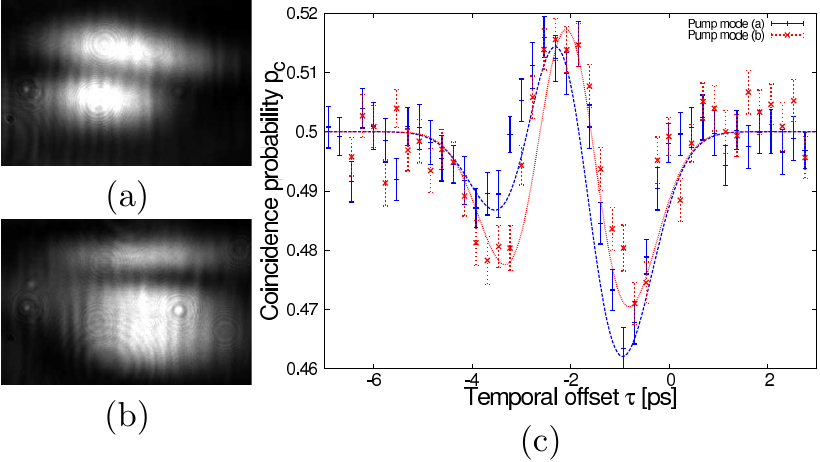}
\end{center}
\caption{(Color online) (c) Experimental results from two HOM measurements with different pump modes (a) and (b); visibility of theory curves (dashed) fitted to match measurement results}\label{results}
\end{center}
\end{figure}

Our measurement results are given in Fig. \ref{results}(c). We observe a beating signature in the coincidence probability \(\prob\), which can be seen as the sum of a bump signature and a wider dip signature. It is in good agreement with the assumption of broadband frequency mode entanglement in our PDC generated bi-photon state: The overlap of the produced photon state with \(\ket{\psi^-}\) is responsible for the central bump, the non-overlapping rest for the dip. We note that in this simple picture, we have truncated our Hilbert space to include only the two lowest order Hermite frequency modes. This result cannot be explained in terms of bi-photonic output states of PDC in bulk crystals if the spectral pump envelope and phasematching functions are modelled as Gaussian functions.

Typically the joint spectral amplitude \(\func{f}{\ws,\wi}\) of a PDC output state \(\ket{\psi_{s,i}}\) is calculated as a product of a spectral distribution of its pump envelope \(\alpha\) and its phasematching function \(\Phi\)
\begin{align}
\func{\alpha}{\ws+\wi}&=\gauss{2\wc-\ws-\wi}{\sigma}\label{pump}\\
\func{\Phi}{\ws,\wi}&=\Sinc{\frac{\dk L}{2}} \etothe{-\imath \frac{\dk L}{2}}\label{phasematching}
\end{align}
where \(\dk\) denotes the phase mismatch, $L$ the crystal length and $\sigma$ the pump bandwidth. We assume the state to be frequency degenerate at \(\wc\), half the pump frequency.
The phasematching function \(\Phi\) is simplified by applying an Gaussian approximation \(\Sinc{x}\approx\etothe{-\gamma x^2}\) with \(\gamma=0.193\).

To construct a joint spectrum in the PDC process which exhibits a significant overlap with the broadband frequency entangled \(\ket{\psi^-}\) state, we require a coherent superposition of two phasematched PDC processes in the Gaussian approximation. Then the joint spectral amplitude shows two distinct maxima
\begin{align}
\func{f^\pm}{\ws,\wi}&=\func{\alpha}{\ws+\wi}\func{\Phi}{\ws\pm\frac{\dw}{2},\wi\mp\frac{\dw}{2}}\label{pmspectrum}\\
\func{F}{\ws,\wi}&=\func{f^+}{\ws,\wi}+r \etothe{\imath\varphi}\func{f^-}{\ws,\wi}\label{spectrum}
\end{align}
Eq. \ref{pmspectrum} represents the spectral distributions symmetrically displaced from \(\degpoint\) by \(\frac{\dw}{\sqrt{2}}\) as illustrated in Fig. \ref{broadbandentangled}(c). The spectral distribution of these superimposed PDC processes with a complex weighting factor \(r \etothe{\imath\varphi}\) is given by Eq. \ref{spectrum}. Comparing Fig. \ref{broadbandentangled}(b) with Fig. \ref{broadbandentangled}(c) already suggests that the overlap of the composite spectral distribution \(\textrm{F}\) with that of \(\ket{\psi^-}\) defined in eq. \ref{bellstate} is non-negligible, which is consistent with our calculations. The HOM interference probability between signal and idler of this state \(\freqint \func{F}{\ws,\wi}\vacstate\) evaluates to
\begin{align}
&\prob=\frac{1}{2} -
\frac
{\etothemfrac{\left(\tau-\tau^-\right)^2}{2 \gamma {\tau^-}^2}}
{2\sqrt{C^+}}
\frac
{\etothemfrac{C^-}{C^+}+\rho\Cos{\dw\,\tau-\varphi}}
{1+\etothe{-C^-} \rho\Cos{\dw \tau^- -\varphi}},\label{masterformula}
\end{align}
where we introduced the expressions
\(\tau^\pm=\left(\dks\pm\dki\right) \frac{L}{2}\),
\(C^- = \frac{\gamma}{2} \dw^2\left(\tau^-\right)^2\),
\(C^+ = 1 + \frac{\gamma}{2} \sigma^2\left(\tau^+\right)^2\),
\(\rho = \frac{2 r}{1+r^2}\),
and \(\dk_{\mu}=\func{k'_p}{2\wc}-\func{k'_\mu}{\wc}\) with \(\mu \in \left\{ s, i \right\}\).

The experimental data, depicted in Fig. \ref{results}(c) matches our theoretical prediction of Eq. \ref{masterformula} closely for a value of \(\dw\) of \uval{1.35}{THz} and \uval{1.5}{THz} respectively. The measurements differ in a slightly different alignment of the spatial pump modes (Fig. \ref{results}(a) and (b)), which was verified with a CCD camera monitoring the discarded pump beam. The reduced visibility of the experiments in comparison to the theory can be understood as a result of experimental imperfections, most prominently a non-Fourier-limited pump laser.

With the HOM fringes clearly visible, we can infer the existence of at least two distinct, coherent spectral maxima of the PDC bi-photon state, modelled by two Gaussian functions \(\mathrm{f^+}\) and \(\mathrm{f^-}\) symmetrically displaced from \(\degpoint\).
An obvious explanation for the spectral structure of our PDC process might seem the \(\mathrm{sinc}\) structure of the phasematching function \(\Phi\), which we had simplified to be Gaussian. However, calculations which use the parameters \(\dw\approx\uval{1.62}{THz}\), \(\rho\approx0.44\), \(\varphi\approx\frac{3}{2}\pi\) to model the \(\mathrm{sinc}\) characteristic, show that for our experimental situation the HOM interference results in a single dip shape. This leaves spatial waveguide mode triples\cite{Banaszek01}, each with a slightly displaced phasematching function due to modal dispersion, as the only possible explanation for our findings. This is also supported by different measurement outcomes for different pump modes.

In conclusion we have demonstrated that a waveguided PDC source is capable of producing broadband frequency mode entangled photon pairs, and that their entanglement affects HOMI experiments. Thus a HOMI signature serves as an witness of entanglement, and allows us to ascertain a significant overlap of the PDC bi-photon with the singlet Bell state \(\ket{\psi^-}\). While this effect surely has to be accounted for in integrated quantum networks using waveguides for state preparation, it also offers the possibility of information coding in the discrete frequency broadband modes for future communication protocols.

\bibliographystyle{ol}

\end{document}